\documentclass{article}
\pdfoutput=1
\usepackage[utf8]{inputenc}
\usepackage[newzealand]{babel}
\usepackage{amsmath}
\usepackage{amsthm}
\usepackage[author-year]{amsrefs}
\usepackage{amssymb}
\usepackage{tikz}
\usetikzlibrary{arrows,calc,shapes}
\usepackage{verbatim}
\usepackage{graphicx}
\usepackage[subrefformat=parens]{subcaption}
\usepackage{booktabs}
\usepackage{aliascnt}
\usepackage[breaklinks=true,pdfborder={0 0 0},pdfdisplaydoctitle=true,
pdfpagelabels=true]{hyperref}
\hypersetup{pdfauthor={Stefan Kranich},pdftitle={GPU-based visualization
of domain-coloured algebraic Riemann surfaces},
pdfkeywords={Riemann surfaces,visualization}}

\usepackage{microtype}

\theoremstyle{plain}
\newtheorem{theorem}{Theorem}[section]

\newcommand{\mynewtheorem}[2]{
    \newaliascnt{#1}{theorem}
    \newtheorem{#1}[#1]{#2}
    \aliascntresetthe{#1}
    \expandafter\def\csname #1autorefname\endcsname{#2}
}

\mynewtheorem{lemma}{Lemma}
\theoremstyle{definition}
\mynewtheorem{algorithm}{Algorithm}
\mynewtheorem{definition}{Definition}
\mynewtheorem{example}{Example}
\mynewtheorem{problem}{Problem}
\theoremstyle{remark}
\mynewtheorem{remark}{Remark}
\makeatletter
\let\c@figure\@undefined
\let\c@table\@undefined
\makeatother
\newaliascnt{figure}{theorem}
\aliascntresetthe{figure}
\numberwithin{figure}{section}
\newaliascnt{table}{theorem}
\aliascntresetthe{table}
\numberwithin{table}{section}

\begin{document}

\title{GPU-based visualization of domain-coloured algebraic Riemann surfaces}
\author{Stefan Kranich\footnote{Zentrum Mathematik (M10), Technische
Universität München, 85747~Garching, Germany; E-mail address:
\url{kranich@ma.tum.de}}}

\maketitle

\begin{abstract}
\noindent We examine an algorithm for the visualization of domain-coloured
Riemann surfaces of plane algebraic curves. The approach faithfully
reproduces the topology and the holomorphic structure of the Riemann
surface.
We discuss how the algorithm can be implemented efficiently in OpenGL with
geometry shaders, and (less efficiently) even in WebGL with multiple
render targets and floating point textures.
While the generation of the surface takes noticeable time in both
implementations, the visualization of a cached Riemann surface mesh is
possible with interactive performance.
This allows us to visually explore otherwise almost unimaginable
mathematical objects.
As examples, we look at the complex square root and the folium of
Descartes. For the folium of Descartes, the visualization reveals
features of the algebraic curve that are not obvious from its equation.
\end{abstract}

\section{Introduction}

\subsection{Mathematical background}
\label{sec:mathematical-background}

The following basic example illustrates what we would like to visualize.

\begin{example}
Let $y$ be the square root of $x$, \[y = \sqrt{x}.\]
If $x$ is a non-negative real number, we typically define $y$ as the
non-negative real number whose square equals $x$, i.e.\ we always choose
the non-negative solution of the equation \begin{equation}
\label{eq:parabola}
y^2 - x = 0
\end{equation}
as $y = \sqrt{x}$. For negative real numbers $x$, no real number $y$ solves
\autoref{eq:parabola}. However, if we define the imaginary unit
$\mathrm{i}$ as a number with the property that $\mathrm{i}^2 = -1$
then the square root of $x$ becomes the purely imaginary number \[y =
\mathrm{i} \sqrt{|x|}.\]
Together, these two conventions yield a continuous square root function
\[\sqrt{\cdot}\colon \mathbb{R} \to \mathbb{C}.\]
For complex numbers $x$, \autoref{eq:parabola} has exactly two complex
solutions (counted with multiplicity), the square roots of $x$.
\end{example}

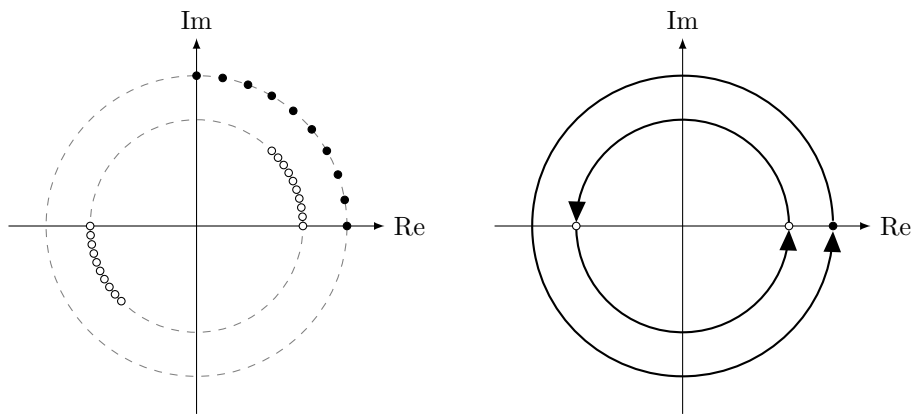
\begin{figure}
\centering
\begin{tikzpicture}
\draw[dashed,gray] (0,0) circle (2);
\draw[dashed,gray] (0,0) circle ({sqrt(2)});
\draw[->,>=latex] (-2.5,0) -- (2.5,0) node[right] {$\operatorname{Re}$};
\draw[->,>=latex] (0,-2.5) -- (0,2.5) node[above] {$\operatorname{Im}$};
\foreach \phi in {0,10,...,90} {
    \filldraw (\phi:2) circle (0.05);
    \draw[fill=white] (0.5*\phi:{sqrt(2)} ) circle (0.05);
    \draw[fill=white] (0.5*\phi+180:{sqrt(2)} ) circle (0.05);
}
\end{tikzpicture}
\hfill
\begin{tikzpicture}
\draw[->,>=latex] (-2.5,0) -- (2.5,0) node[right] {$\operatorname{Re}$};
\draw[->,>=latex] (0,-2.5) -- (0,2.5) node[above] {$\operatorname{Im}$};
\filldraw (0:2) circle (0.05);
\draw[fill=white] (0:{sqrt(2)} ) circle (0.05);
\draw[fill=white] (180:{sqrt(2)} ) circle (0.05);
\draw[thick,->,>=triangle 45] (2:{sqrt(2)} ) arc (2:178:{sqrt(2)} );
\draw[thick,->,>=triangle 45] (182:{sqrt(2)} ) arc (182:358:{sqrt(2)} );
\draw[thick,->,>=triangle 45] (2:2) arc (2:358:2);
\end{tikzpicture}
\caption{When a complex number (black points) runs along a circle centred
at the origin of the complex plane, its square roots (white points) move
at half the angular velocity (left image). After a full turn of $x$,
the square roots have interchanged positions. The real part of the square
root that initially had positive real part has become negative, and vice versa
(right image).}
\label{fig:square-root-angular-velocity}
\end{figure}

\noindent We have seen that, for real numbers $x$, we can choose one solution
of~\autoref{eq:parabola} for the square root and obtain a square root
function that is continuous over the real numbers.
In contrast, we cannot for every complex number $x$ choose one solution
of~\autoref{eq:parabola} so that we obtain a square root function
that is continuous over the complex numbers:
If we plot the two solutions of~\autoref{eq:parabola} as $x$ runs along
a circle centred at the origin of the complex plane, we observe that $y$
moves at half the angular velocity of $x$
(see~\autoref{fig:square-root-angular-velocity}).
When $x$ completes one full circle and reaches its initial position again,
the square roots have interchanged signs. Therefore, a discontinuity
occurs when $x$ returns to its initial position after one full turn and
the square root jumps back to its initial position.

Note that by choosing the values of the square root in a different manner,
or, equivalently, letting $x$ start at at a different position, we can
move the discontinuity to an arbitrary position on the circle.
Moreover, note that there is (at least) one discontinuity on any circle
of any radius centred at the origin.

In order to define the principal branch of the complex square root
function, we usually align the discontinuities along the negative real
axis, the canonical branch cut of the complex square root function,
and choose those values that on the real axis agree with the square
root over the real numbers.

Alternatively, we can extend the domain of the complex square root to
make it a single-valued and continuous function. To that end, we take
two copies of the extended complex plane and slit them along the negative real
axis. On the first copy, we choose the solution of~\autoref{eq:parabola}
with non-negative real part as the complex square root of $x$; on the second
copy, we choose the other solution.
We glue the upper side (lower side) of the slit of the first copy to
the lower side (upper side) of the slit of the second copy and obtain
a Riemann surface of the complex square root. (In three dimensions,
this is not possible without self-intersections.)

\begin{remark}
On the Riemann surface, the complex square root is single-valued and
continuous. It is even analytic except at the origin and at
infinity, which are exactly the points where the two solutions of
\autoref{eq:parabola} coincide.
\end{remark}

\begin{remark}
The branch cut is not a special curve of the Riemann surface. When we glue
the Riemann surface together, the branch cut becomes a curve like every
other curve on the Riemann surface. If we had used a different curve
between the origin and infinity as branch cut, we would have obtained
the same result.
\end{remark}

\begin{remark}
\autoref{eq:parabola} describes a parabola. We can proceed analogously
to obtain Riemann surfaces for other plane algebraic curves.
\end{remark}

\subsection{Previous work}

Probably the most common approach for visualization of functions is to
plot a function graph.
However, for a complex function \[g\colon \mathbb{C} \to \mathbb{C},\] the
function graph \[\{(z, g(z)) \mid z \in \mathbb{C}\} \subset \mathbb{C}
\times \mathbb{C} \simeq \mathbb{R}^2 \times \mathbb{R}^2\] is a (real)
two-dimensional surface in (real) four-dimensional space.

One way to visualize a four-dimensional object is to plot several
two- or three-dimensional slices. This approach seems less useful for
understanding the overall structure of the object.

Another traditional method to visualize complex functions is domain
colouring.
The principle of domain colouring is to colour every point in the domain
of a function with the colour of its function value in a reference image.
If we choose the reference image wisely, a lot of information about the
complex function can be read off from the resulting two-dimensional
image (see e.g.~\cite{PoelkePolthier2012} and~\cite{Wegert2012}).
The idea of lifting domain colouring to Riemann surfaces is due to
\ocite{PoelkePolthier2009}.

We can interpret a Riemann surface of a plane algebraic curve \begin{equation}
\mathcal{C}\colon f(x,y) = 0
\end{equation} as a function graph of a multivalued complex function,
which maps every $x$ to multiple values of $y$.
If $f(x,y)$ is a polynomial of degree $n$ in $y$, there are
exactly $n$ values of $y$ for every value of $x$ that satisfy $f(x,y)
= 0$ (counted with multiplicity).
Every such pair $(x,y)$ corresponds to a point on the Riemann surface. In
other words, the Riemann surface is an $n$-fold cover of the complex plane.

Let $\pi\colon (x,y) \mapsto x$ denote a projection function on the
Riemann surface. Then the values of $y$ at $x$ correspond to the elements
of the fibre $\pi^{-1}(x)$.
The situation is analogous to function graphs of single-valued functions
from the real numbers (or the real plane) to the real numbers, where one
function value lies above every point in the domain.

We can transfer the Riemann surface from (real) four-dimensional space
into (real) three-dimensional space by introducing a height function
$H\colon \mathbb{C} \to \mathbb{R}$. We typically use the real part as
a height function. We plot the surface
\[\{(\operatorname{Re} x, \operatorname{Im} x, H (y)) \mid x, y \in
\mathbb{C}\colon f (x, y) = 0\}\]
and use domain colouring to represent the value of $y$ at every point
of the surface.

In practice, we want to generate a triangle mesh that approximates the
Riemann surface as the graph of a multivalued function over a triangulated
domain in the complex plane.
The Riemann surface mesh approximates the continuous Riemann
surface in the following sense: The $y$-values at the vertices of
a triangle of the Riemann surface mesh result from each other under
analytic continuation along the edges of the underlying triangle in the
triangulated domain.
If $f(x,y) = 0$ is a polynomial of degree $n$ in $y$ there are $n$
values of $y$ above every vertex $x$ of the triangulated domain.
Hence, we have to determine which of the $3 n$ values of $y$ above
a triangle in the triangulated domain form triangles of the Riemann
surface mesh.
A wrong combination of values of $y$ to triangles might for example occur
due to discontinuity if we used the principal branch of the square root
function for the computation of $y$. This would produce artefacts in
the visualization for which there is no mathematical justification.

For the generation of such a Riemann surface mesh, previous algorithms
have solved systems of differential equations~\cites{Trott2008,
NieserPoelkePolthier2010} or explicitly identified and analyzed
branch cuts to remove discontinuities~\citelist{\cite{Kranich2012}
\cite{Wegert2012}*{Section~7.6}}.

In the next section, we discuss an algorithm based on a different
idea: We can exploit that $y(x)$ is continuous almost everywhere on
the Riemann surface and therefore, if $x$ changes little, so does $y(x)$.

\section{Algorithms}

In this section, we describe algorithms for generating and visualizing
domain-coloured Riemann surface meshes of plane algebraic curves.
Let \begin{equation}
\label{eq:plane-algebraic-curve}
\mathcal{C}\colon f (x, y) = 0
\end{equation} be a complex plane algebraic curve. In particular, let
$f$ be a polynomial with complex coefficients of degree $n$ in $y$.
Moreover let $U \subset \mathbb{C}$ be a triangulated domain in the
complex plane. (In practice, $U$ is typically rectangular.)

We want to generate a Riemann surface mesh of $\mathcal{C}$. The mesh
discretizes a part of a (real) two-dimensional surface in (real)
four-dimensional space. We can visualize it using a height function and
domain colouring, as described in the previous section.

We obtain a Riemann surface mesh of $\mathcal{C}$ as a graph of
the multivalued function induced by~\autoref{eq:plane-algebraic-curve},
which maps every value of $x$ in $U$ to $n$ values of $y$ such that $f
(x, y) = 0$.

For every triangle in $U$, we thus obtain $n$ values of $y$
at each of its three vertices. The problem is to determine whether,
and if so, how, the $3 n$ values of $y$ can be combined to
form triangles of the Riemann surface mesh. The resulting triangles
should be consistent with the fact that $y$ as a
function of $x$ is analytic almost everywhere on the
Riemann surface. This is impossible if the triangle in $U$ contains
a ramification point of $y(x)$. In this case, we subdivide the triangle
to obtain smaller triangles mostly free of ramification points. Otherwise,
the triangles of the Riemann surface mesh are uniquely determined by
analytic continuation of $y(x)$ along the edges of the triangle in $U$.

In order to find these triangles of the Riemann surface mesh, we use
the following idea:
Consider a triangle $\triangle x_1 x_2 x_3$ in $U$ that is free of
ramification points of $y(x)$.
Under this assumption, $y(x)$ is continuous on those parts of the Riemann
surface that lie above $\triangle x_1 x_2 x_3$.
Hence, for every
$\varepsilon > 0$ there exists $\delta > 0$ such that $|y(x_1) - y(x_2)|
< \varepsilon$ for all $x_1, x_2$ with $|x_1 - x_2| < \delta$.
If $\varepsilon$ is half the minimum distance between the $n$ values of
$y(x)$ at $x_1$ and $|x_1 - x_2|$ is smaller than the corresponding $\delta$,
then the values of $y(x)$ at $x_2$ are closer to the corresponding values
of $y(x)$ at $x_1$ than to any other value of $y(x)$ at $x_1$.

In other words, if triangle $\triangle x_1 x_2 x_3$ is small enough, we can
combine the values of $y$ at its vertices to triangles of the Riemann
surface mesh based on proximity: Among the $3 n$ values of $y$ at the
vertices of triangle $\triangle x_1 x_2 x_3$, every three values of $y$
closest to each other form a triangle of the Riemann surface mesh.

We can algorithmically compute a $\delta > 0$ as above using the epsilon-delta
bound for plane algebraic curves of~\autoref{thm:epsilon-delta-bound}.
\autoref{thm:epsilon-delta-bound} is of essential
importance for our approach. Our approach only works because
\autoref{thm:epsilon-delta-bound} provides us with a reliable bound
computable as a function of $x$ that depends only on a few constants
derived from the coefficients of $f(x,y)$.

If triangle $\triangle x_1 x_2 x_3$ is not small enough to correctly
combine the values of $y$ at its vertices based on proximity, we subdivide
the triangle.

In summary, we obtain the following algorithm:

\begin{algorithm}[Generation of a Riemann surface mesh]
\label{alg:riemann-surface-mesh}
Let $U \subset \mathbb{C}$ be a triangulated domain in the complex plane.
Let \[\mathcal{C}\colon f (x, y) = 0\] be a complex plane algebraic
curve and $f (x, y)$ a polynomial of degree $n$ in $y$.
We prescribe a maximal subdivision depth (as a maximal number of
iterations or as a minimal edge length).
\begin{enumerate}
\item Compute the global ingredients of the epsilon-delta bound of
\autoref{thm:epsilon-delta-bound} for $y (x)$.
\item For every triangle $\triangle x_1 x_2 x_3$ in $U$:
\begin{enumerate}
\item Compute the $3 n$ values of $y(x)$ at $x_1, x_2, x_3$, \[\{y_k
(x_j) \mid f (x_j, y_k (x_j)) = 0,\ j = 1, 2, 3,\ k = 1, 2, \dots, n\}.\]
\item Compute half the minimum distance between the values of $y(x)$
at each of the vertices of $\triangle x_1 x_2 x_3$, \[\varepsilon (x_j) =
\tfrac{1}{2} \min_{k \neq l} |y_k(x_j) - y_l(x_j)|, \quad j = 1, 2, 3.\]
\item Compute $\delta (x_j)$ by the epsilon-delta bound of
\autoref{thm:epsilon-delta-bound} so that \[|y(x_j) - y(x)| <
\varepsilon (x_j), \text{ if } |x_j - x| < \delta (x_j), \quad j = 1, 2, 3.\]
\item Determine which of the edges of $\triangle x_1 x_2 x_3$ are longer
than the minimum of the $\delta (x_j)$ at their endpoints and must
be subdivided.
\item Select the right adaptive refinement pattern
(see~\autoref{fig:adaptive-refinement-patterns}) and subdivide $\triangle
x_1 x_2 x_3$ accordingly.
\end{enumerate}
\item Repeat step 2 until the maximal subdivision depth is reached.
\item Discard every triangle in $U$ with an edge longer than the minimum
of the $\delta (x_j)$ at its endpoints.
\item For every triangle $\triangle x_1 x_2 x_3$ in $U$, combine the
values of $y(x)$ at its vertices to triangles of the Riemann surface
mesh based on proximity. More formally, the triangles added to the Riemann
surface mesh comprise the vertices
\begin{gather*}
(x_1, y_k(x_1)),\\
(x_2, \operatorname*{argmin}_{y_l (x_2)} |y_k (x_1) - y_l(x_2)|),\\
(x_3, \operatorname*{argmin}_{y_l (x_3)} |y_k(x_1) - y_l(x_3)|),
\end{gather*} for $k = 1, 2, \dots, n$.
\item Output the Riemann surface mesh and stop.
\end{enumerate}
\end{algorithm}

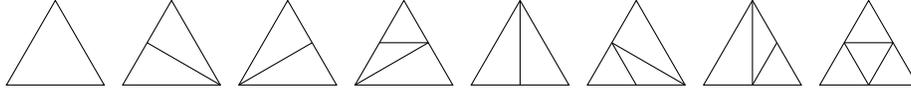
\begin{figure}
\centering
\begin{tikzpicture}[scale=0.75]
\coordinate (A) at (210:1);
\coordinate (B) at (330:1);
\coordinate (C) at (90:1);
\draw (A) -- (B) -- (C) -- cycle;
\end{tikzpicture}
\hfill
\begin{tikzpicture}
\draw (A) -- (B) -- (C) -- cycle;
\draw (B) -- ($(C)!0.5!(A)$);
\end{tikzpicture}
\hfill
\begin{tikzpicture}
\draw (A) -- (B) -- (C) -- cycle;
\draw (A) -- ($(B)!0.5!(C)$);
\end{tikzpicture}
\hfill
\begin{tikzpicture}
\draw (A) -- (B) -- (C) -- cycle;
\draw (A) -- ($(B)!0.5!(C)$);
\draw ($(C)!0.5!(A)$) -- ($(B)!0.5!(C)$);
\end{tikzpicture}
\hfill
\begin{tikzpicture}
\draw (A) -- (B) -- (C) -- cycle;
\draw (C) -- ($(A)!0.5!(B)$);
\end{tikzpicture}
\hfill
\begin{tikzpicture}
\draw (A) -- (B) -- (C) -- cycle;
\draw ($(A)!0.5!(B)$) -- ($(C)!0.5!(A)$);
\draw (B) -- ($(C)!0.5!(A)$);
\end{tikzpicture}
\hfill
\begin{tikzpicture}
\draw (A) -- (B) -- (C) -- cycle;
\draw (C) -- ($(A)!0.5!(B)$);
\draw ($(A)!0.5!(B)$) -- ($(B)!0.5!(C)$);
\end{tikzpicture}
\hfill
\begin{tikzpicture}
\draw (A) -- (B) -- (C) -- cycle;
\draw ($(A)!0.5!(B)$) -- ($(B)!0.5!(C)$);
\draw ($(B)!0.5!(C)$) -- ($(C)!0.5!(A)$);
\draw ($(C)!0.5!(A)$) -- ($(A)!0.5!(B)$);
\end{tikzpicture}
\caption{Adaptive refinement patterns used in
\autoref{alg:riemann-surface-mesh}}
\label{fig:adaptive-refinement-patterns}
\end{figure}

\begin{remark}
By construction, \autoref{alg:riemann-surface-mesh} generates a Riemann
surface mesh that is consistent with the analytic structure of the
Riemann surface of $\mathcal{C}$.
\end{remark}

\begin{remark}
The adaptive refinement patterns used for the subdivision of triangles,
whose edges are too long, produce a watertight subdivision.
\end{remark}

\begin{remark}
Step 4 of~\autoref{alg:riemann-surface-mesh} produces holes around the
ramification points of $y(x)$. We can make these holes very small if we
choose the maximal subdivision depth appropriately.
\end{remark}

\noindent For the visualization of a Riemann surface mesh, we use the
following algorithm:

\begin{algorithm}[Visualization of a Riemann surface mesh]
\label{alg:visualization}
Let a Riemann surface mesh and a domain colouring reference image be
given. We choose a height function $H\colon \mathbb{C} \to \mathbb{R}$ to
transform a point on the Riemann surface mesh from (real) four-dimensional
space to a point in (real) three-dimensional space, \[(x, y(x)) \mapsto
(\operatorname{Re} x, \operatorname{Im} x, H (y(x))).\]
\begin{enumerate}
\item Draw the mesh that results from transforming every vertex of the
Riemann surface mesh as above.
\item Interpolate the value of $y(x)$ on the transformed mesh.
\item Assign to every point on the transformed mesh the colour in the
reference image of the value that $y(x)$ attains at that point on the
transformed mesh.
\end{enumerate}
\end{algorithm}

\begin{remark}
If we choose the real (or imaginary) part of $y(x)$ as a height
function, the transformation from (real) four-dimensional to (real)
three-dimensional space becomes a projection.
\end{remark}

\begin{remark}
Using the real part of $y(x)$ as a height function has the advantage
that the visualization then contains the image of $\mathcal{C}$ interpreted
as a real plane algebraic curve. It is the intersection of the
visualization of the Riemann surface mesh in (real) three-dimensional
space with the $\operatorname{Re} x$-$\operatorname{Re} y$-plane (the
$xz$-plane, if we label the coordinate axes of real three-dimensional
space such that the $x$-axis points to the right and the $z$-axis points
upwards).
\end{remark}

\begin{remark}
The computation of the Riemann surface mesh by
\autoref{alg:riemann-surface-mesh} is independent of the choice of
height function used for its visualization.
\end{remark}

\section{Implementation}

In this section, we discuss how~\autoref{alg:riemann-surface-mesh}
and~\autoref{alg:visualization} can be implemented using OpenGL and
WebGL.\@ Since WebGL targets a much wider range of devices, its API is
more limited than that of OpenGL.\@ Consequently, our implementation using
WebGL differs substantially from our implementation using OpenGL.\@ Before
we discuss each setup separately, let us talk about what they have in common.

The main part of our programs is written in shading language (GLSL for
OpenGL and ESSL for WebGL) and runs on the GPU.\@
We use the CPU to compute the global ingredients for the epsilon-delta
bound, to generate shading language code that computes the epsilon-delta
bound for $\mathcal{C}\colon f(x,y) = 0$ as a function of $x$, and to
generate a coarse triangulation of the input domain.

The implementations in OpenGL and WebGL share some shading language
code. Since there is no native type for complex numbers, we represent them
using two-dimensional floating point vectors. Common routines include
complex arithmetic, numerical root-finding algorithms, the computation
of the epsilon-delta bound, and domain colouring.

The implementation of complex arithmetic is straightforward and we shall
not go into detail about it.

We need numerical root-finding algorithms to approximate
roots of polynomials in order to compute values of
$y(x)$ (and to compute the global ingredients of the
epsilon-delta bound). For instance, Laguerre's
method~\cite{PressTeukolskyVetterlingFlannery2007}*{Section~9.5.1} and
deflation~\cite{PressTeukolskyVetterlingFlannery2007}*{Section~9.5.3}
or Weierstraß--Durand--Kerner
method~\cites{Weierstrass1891,Durand1960,Kerner1966} are well-suited. The
latter may be a little easier to implement in shading language (due to
the absence of variable-length arrays).

For the computation of the epsilon-delta bound, we use the following theorem:
\begin{theorem}[cf.~\cite{Kranich2015}*{Theorem~2.1}]
\label{thm:epsilon-delta-bound}
Let $\mathcal{C}\colon f(x, y) = 0$ be a complex plane algebraic curve,
where \[f(x, y) = \sum\limits_{k = 0}^n a_k(x) y^{n - k}\] is a polynomial
of degree $n$ in $y$ whose coefficients $a_k(x)$ are polynomials in $x$ of the form
\[a_k(x) = \sum_{l = 0}^{m_k} a_{kl} x^{m_k - l}.\]
Let $x_1 \in \mathbb{C}$ be a point
in the complex plane at which neither the leading coefficient $a_0(x)$
nor the discriminant of $f(x,y)$ w.r.t.\ $y$ vanish.
Then for every $\varepsilon > 0$, we can algorithmically compute $\delta >
0$ such that \[|y_j(x_1) - y_j(x_2)| < \varepsilon\] for all holomorphic
functions $y_j(x)$, $j = 1, 2, \dots, n$, which satisfy $f(x, y_j(x)) = 0$
in a neighbourhood of $x_1$ and for all $x_2$ with $|x_1 - x_2| < \delta$.

We obtain
\begin{equation}
\label{eq:epsilon-delta-bound}
\delta = \frac{\rho \left(\sqrt{{(\rho Y - \varepsilon)}^2 + 4 \varepsilon
M} - (\rho Y + \varepsilon)\right)}{2 (M - \rho Y)},
\end{equation}
where
\begin{gather*}
\rho < \min\{|x_1 - x| \colon a_0(x) \cdot \Delta_y(f(x,y))(x) = 0\},\\
Y:= \max\limits_j \left|\dfrac{f_x(x_1, y_j(x_1))}{f_y(x_1,
y_j(x_1))}\right|,\quad
M := 2 \max\limits_k
{\left(\frac{\tilde{a}_k}{\tilde{a}_0}\right)}^{\frac{1}{k}},\\
\tilde{a}_0 := |a_{00}| \prod_{l = 1}^{m_0} (|\bar{x}_l - x_1| - \rho) > 0,\quad
\tilde{a}_k := \sum_{l = 0}^{m_k} |a_{kl}| {(|x_1| + |\rho|)}^{n-l},
\end{gather*}
$\Delta_y(f(x,y))(x)$ denotes the $y$-discriminant of $f(x,y)$, and
$\bar{x}_l$, $l = 1, 2, \dots, m_0$, are the zeros of $a_0 (x)$.
\end{theorem}

Note that the computation is parallelizable since the epsilon-delta
bound can be implemented as a function of $x$ that depends on only a
few constants derived from the coefficients of $f(x,y)$.

Instead of computing texture coordinates, which would depend on the range of
$y(x)$ on the input domain, we generate the domain colouring procedurally
on-the-fly. To that end, we use a variation of the enhanced phase portrait
colour scheme of~\cite{Wegert2012}*{Section~2.5}. The reference image
is shown in~\autoref{fig:domain-colouring}. We discuss the colour scheme
in~\autoref{sec:reference-image}.

The main difference between the implementations in OpenGL and WebGL is
how the common routines can be combined to realize
\autoref{alg:riemann-surface-mesh} and~\autoref{alg:visualization}.

\subsection{An implementation in OpenGL}

\subsubsection{Implementation of~\autoref{alg:riemann-surface-mesh} in OpenGL}

Our implementation of~\autoref{alg:riemann-surface-mesh} in OpenGL
comprises three GLSL programs, for initialization, subdivision, and
assembly of the Riemann surface mesh. We cache the output of each program
using transform feedback and feed it back to the next program.

The initialization program consists only of a vertex shader, which operates on
the vertices of the triangulated input domain. For every vertex $x$, we
compute $y_k (x)$, $k = 1, 2, \dots, n$, and $\delta (x)$.

After initialization, we run the subdivision program. The program consists of
a pass-through vertex shader and a geometry shader. The geometry shader
operates on the triangles of the triangulated input domain or of its
last subdivision, respectively. We have access to the values of $x_j$,
$\delta (x_j)$, and $y_k (x_j)$, $k = 1, 2, \dots, n$, $j = 1, 2, 3$,
at the vertices of each triangle $\triangle x_1 x_2 x_3$.
We determine which edges of triangle $\triangle x_1 x_2 x_3$ are longer
than the minimum of the $\delta (x_j)$ at their endpoints. In order
to subdivide these edges, we compute their midpoints $x$, and $\delta
(x)$ and $y_k (x)$, $k = 1, 2, \dots, n$, at the
midpoints. We use the appropriate adaptive refine pattern of
\autoref{fig:adaptive-refinement-patterns} and output between one and
four triangles for every input triangle. In doing so, we reuse previously
computed values rather than recomputing them.
We run the subdivision program iteratively until we reach the prescribed
maximal subdivision depth.

The assembly program consists of a pass-through vertex shader and a
geometry shader. The geometry shader operates on the triangles of the
adaptively subdivided input domain. We again have access to the values
of $x_j$, $\delta (x_j)$, and $y_k (x_j)$, $k = 1, 2, \dots, n$, $j =
1, 2, 3$, at the vertices of each triangle $\triangle x_1 x_2 x_3$.
For every triangle $\triangle x_1 x_2 x_3$, we test whether one of
its edges is longer than the minimum of the $\delta (x_j)$ at its
endpoints. In this case, we discard the triangle.
Otherwise, we determine the triangles of the Riemann surface mesh by
proximity (see~\autoref{alg:riemann-surface-mesh}, step~5) and output
these $n$ triangles.

We also cache the assembled Riemann surface mesh using transform
feedback so that we can pass it as input to our implementation of the
visualization algorithm (\autoref{alg:visualization}).

\subsubsection{Implementation of~\autoref{alg:visualization} in OpenGL}

Our implementation of~\autoref{alg:visualization} in OpenGL consists of
one GLSL program with a vertex and a fragment shader.

The vertex shader operates on the vertices of a Riemann surface mesh
generated by our implementation of~\autoref{alg:riemann-surface-mesh}.
We apply height function $H\colon \mathbb{C} \to \mathbb{R}$ to map each
(real) four-dimensional vertex \[(\operatorname{Re} x, \operatorname{Im}
x, \operatorname{Re} y (x), \operatorname{Im} y (x))\] to a (real)
three-dimensional vertex \[(\operatorname{Re} x, \operatorname{Im} x,
H (y (x))).\]
We homogenize the coordinates of this vertex and transform them using
the model-view-projection matrix.
We pass $y (x)$ as a varying variable to the fragment shader.

The fragment shader operates on the interpolated value of $y (x)$ at a
fragment of a pixel of the output device.
We compute the colour of $y (x)$ according to our domain colouring
reference image.

\subsubsection{Remarks}

Using our implementation, the generation of a Riemann surface mesh takes
little but noticeable time. The bottlenecks of the implementation are
numerical root-finding and iterative subdivision. However, if we use
transform feedback to cache the Riemann surface mesh and pass it to the
implementation of the visualization algorithm, we obtain interactive
performance.

\begin{figure}[ht!]
\centering
\includegraphics[width=\linewidth]{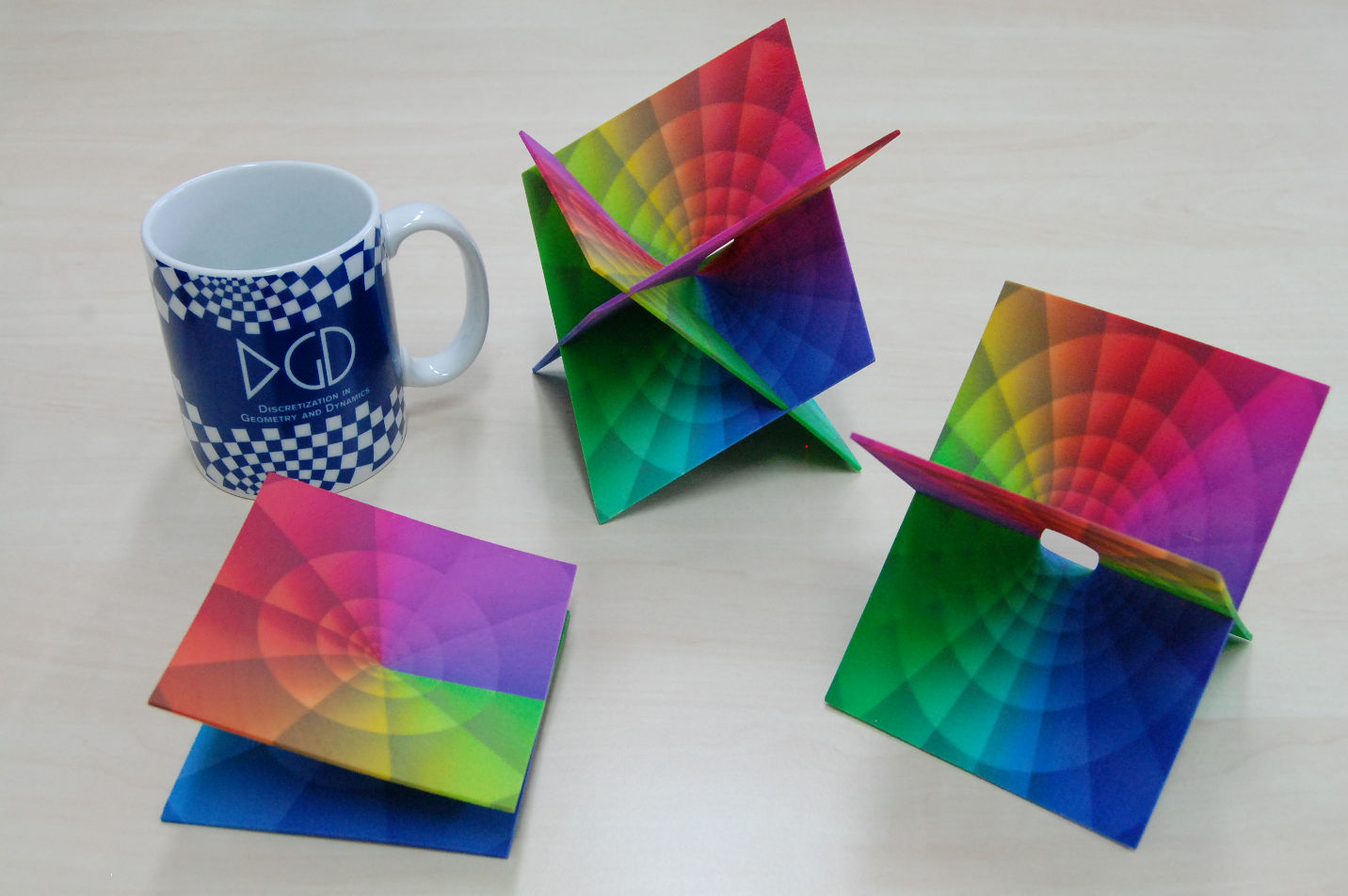}
\caption{3D-printed models of domain-coloured Riemann surfaces of square
root, folium of Descartes, and unit circle (clockwise).
The merchandise coffee mug of DFG Collaborative Research Center TRR 109,
``Discretization in Geometry and Dynamics'', is included in the picture
as an indicator for the size of the models.}
\label{fig:3d-printed-models}
\end{figure}

Another advantage of using transform feedback to cache the Riemann
surface mesh is that we can easily export the data. If we additionally compute
texture coordinates and a high-resolution reference image, we can even print
our visualization using a full colour 3D printer (see
\autoref{fig:3d-printed-models}).

\subsection{An implementation in WebGL}

In order to support a wider range of devices, the WebGL API is much more
limited than the OpenGL API.\@ Particularly, in WebGL, geometry shaders
and transform feedback are currently unavailable. (The WebGL 2 draft
includes transform feedback and compute shaders.)
Therefore our implementation in WebGL differs substantially from our
implementation in OpenGL.\@

\subsubsection{How to replace transform feedback}

Instead of transform feedback, our implementation in WebGL uses floating
point textures (specified in the \texttt{OES\_texture\_float}
extension) and multiple render targets (specified in the
\texttt{WEBGL\_draw\_buffers} extension).
I do not claim originality of this approach. It is commonly used for
running simulations on the GPU.\@ The original idea may be due
to~\cite{Crane2007}.
We number the vertices of every mesh consecutively and pass this number
(index) to the vertex shaders along with the other attributes. In
particular, vertices that are shared among several triangles must be
duplicated and numbered separately.
Hence, we assume that every triangle appears as three consecutive vertices
in array buffer storage (triangle soup).
We use floating point textures essentially as we would arrays of floats,
indexed by vertex number.
We store values corresponding to the $k$-th vertex in the $k$-th pixel
of a texture.
We can store up to four floats per pixel of a floating point texture, namely
one float each in the red, green, blue, and alpha channel.
If we need to store more than four floats, we use multiple render targets 
which allows us to colour the same pixel of several textures simultaneously.

We want to store values we compute for a vertex in textures (`transform'
in transform feedback). To that end, we bind the array buffers and draw
the contents as points (as opposed to triangles). In the vertex shader,
we compute the positions of the point with index $k$ (in normalized
device coordinates) so that it is rasterized as the $k$-th pixel of the
render target textures. Recall that normalized device coordinates range
in ${\left[-1,1\right]}^3$. For example, if $h$
and $w$ denote the height and width of the textures in pixels, we assign
to the point with index $k$ the position
\[\left(\frac{2 \cdot (k \text{ mod } w) + 1}{w} - 1,
\frac{2 \cdot \lfloor k / w\rfloor + 1}{h}- 1, 0\right).\]
In the fragment shader, we compute the values to be stored and assign
them as output colours in a specific order (as we later want to retrieve them).

We want to read stored values for a vertex from textures (`feedback'
in transform feedback). To that end, we bind the textures and an array
buffer containing a range of vertex numbers (indices). We draw the
contents of the array buffer as points or triangles (depending on whether
we want to send the output to different textures or to the screen).
In the vertex shader, we compute texture coordinates for the point with
index $k$ which allow us to lookup the $k$-th pixel from the textures.
Recall that texture coordinates range in ${\left[0,1\right]}^2$.
For a texture of height $h$ and width $w$, we compute the texture coordinates
\[\left(\frac{(k \text{ mod } w) + 0.5}{w},\frac{(k \text{ mod } w)
+ 0.5}{h}\right).\]
Adding $0.5$ in the numerators accounts for the fact that we want to obtain
coordinates for the centre of a pixel in order to avoid interpolation
with adjacent pixels.
We pass the texture coordinates to the fragment shader, where we can
use them to perform a texture lookup.

In order to access data of a whole triangle (as in geometry shaders),
we can, in the vertex shader, determine the indices of the other vertices
of the triangle. For example, the point with index $k$ is part of the
triangle whose vertices have indices \[k - (k \text{ mod } 3),\ k -
(k \text { mod } 3) + 1, \text{ and } k - (k \text { mod } 3) + 2.\]
We compute normalized device coordinates or texture coordinates for all
three indices and pass them to the fragment shader, together with the
index of the triangle vertex currently under consideration.

\subsubsection{How to replace geometry shaders}

We replace the geometry shader of the subdivision program
of our implementation in OpenGL using a variation of a method proposed by
\ocite{BoubekeurSchlick2008}. The method works as follows:
We precompute all adaptive refinement patterns up to a
certain subdivision depth, in our case eight adaptive refinement patterns
up to depth one (see~\autoref{fig:adaptive-refinement-patterns}).
We use barycentric coordinates to store the positions of the triangle
vertices of each refinement pattern in an array buffer.
Using array buffers of different lengths allows us to achieve
variable-length output, as with geometry shaders.
For every triangle of a coarse input mesh,
we draw the triangles in the array buffer of the appropriate adaptive
refinement pattern. We use the vertex positions of the input triangle
(read from a texture or from uniform variables) and the barycentric
coordinates of the triangles of the adaptive refinement pattern to
compute the vertex positions of the output triangle.

We can combine this method with floating point texture and multiple render
targets as outlined above, if we number the vertices of each adaptive
refinement pattern consecutively and store those indices together with
the barycentric coordinates. We pass an offset as a uniform variable to
the vertex shader that needs to be added to the indices. We draw the
adaptive refinement pattern and increment the offset by the number of
vertices in the adaptive refinement pattern.

The geometry shader of the assembly program has fixed-length output. It
generates exactly $n$ triangles of the Riemann surface mesh per triangle
of the (subdivided) input mesh. We can replace it with $n$ invocations
of a vertex shader, one for every sheet of the Riemann surface mesh. We
pass the number of the current sheet to the vertex shader as a uniform
variable.

\subsubsection{Remarks}

We cannot expect our WebGL implementation to reach the same performance as our
OpenGL implementation. In the subdivision program, since we draw a
different adaptive refinement pattern for every triangle of the input
mesh, we lose parallelism. Consequently, subdivision in WebGL is much
slower than its OpenGL counterpart.
However, if we cache the assembled Riemann surface mesh (in textures)
and pass it to our implementation of the visualization algorithm, we
can still achieve interactive performance.

\section{Examples}

In this section, we discuss domain-coloured Riemann surface meshes for
the complex square root function and for the folium of Descartes.
Before that, let us explain our domain colouring reference image so that
we can interpret the domain-coloured Riemann surface meshes.

\subsection{Domain colouring reference image}
\label{sec:reference-image}

Recall that the basic idea of domain colouring is the following: If we
want to visualize a complex function \[f\colon K \subset \mathbb{C} \to
\mathbb{C},\] we face the problem that its graph is real four-dimensional.
However, we can visualize the behaviour of the function by colouring
every point in its domain with the colour of the function value at that
point in a reference image.
The reference image is the domain colouring of the complex identity function.

Depending on what reference image we choose, we can read off various
properties of a function from its domain colouring. For
an overview of different colour schemes, we refer
to~\cites{PoelkePolthier2012,Wegert2012}.

As our reference image, we use a variation of the enhanced phase portrait
colour scheme of~\cite{Wegert2012}*{Section~2.5}.
The reference image is best described using polar coordinates \[r
\mathrm{e}^{\mathrm{i} \varphi} = r (\cos\varphi + \mathrm{i}
\sin\varphi)\] of a complex number with \emph{modulus} $r > 0$ and
\emph{phase} $\varphi \in \left[0,2\pi\right)$.

\begin{figure}[ht]
\centering
\setcounter{subfigure}{0}
\begin{subfigure}[b]{0.32\linewidth}
\includegraphics[width=\linewidth]{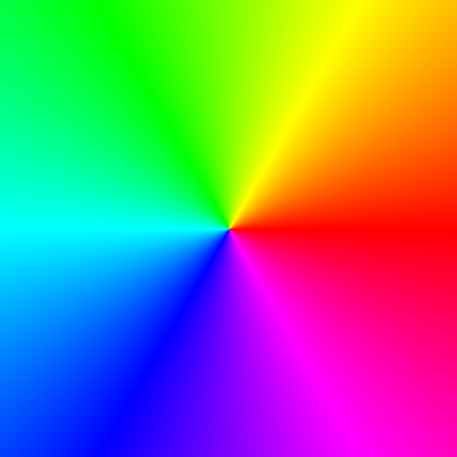}
\caption{}\label{fig:domain-colouring-hue}
\end{subfigure}
\begin{subfigure}[b]{0.32\linewidth}
\includegraphics[width=\linewidth]{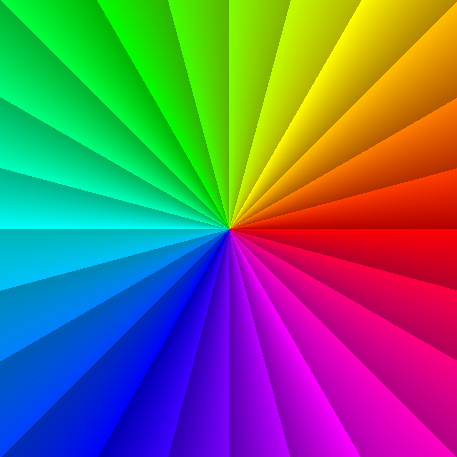}
\caption{}\label{fig:domain-colouring-blackp}
\end{subfigure}
\begin{subfigure}[b]{0.32\linewidth}
\includegraphics[width=\linewidth]{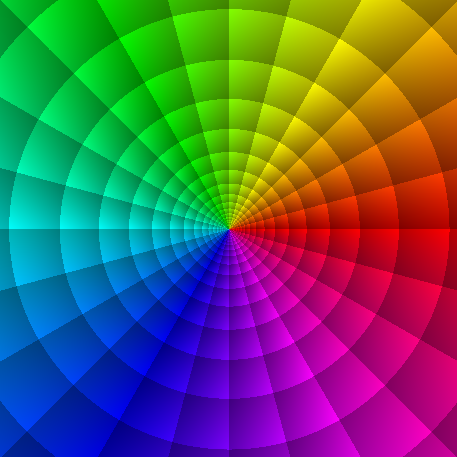}
\caption{}\label{fig:domain-colouring-reference-image}
\end{subfigure}
\caption{Composition of domain colouring reference image: If
we \subref{fig:domain-colouring-hue} represent phase by hue,
\subref{fig:domain-colouring-blackp} add contour lines of phase
and \subref{fig:domain-colouring-reference-image} add contour lines of modulus,
we obtain our domain colouring reference image, a variation of the
enhanced phase portrait colour scheme of~\cite{Wegert2012}*{Section~2.5}.}
\label{fig:domain-colouring}
\end{figure}

Firstly, we encode the phase at any point in the domain as the hue of
its colour (in HSI colour space). In a square with side length $10$
centred at the origin, we thus obtain the colour wheel shown
in~\autoref{fig:domain-colouring-hue}.
As the phase changes from $0$ to $2 \pi$, we obtain every colour
of the rainbow. Positive real numbers, which have phase $0$, are coloured
in pure red. Negative real numbers, which have phase $\pi$, are coloured in
cyan. Purely imaginary numbers do not have such distinctive colours. (This
can be fixed using the NIST continuous phase mapping, which scales the
phase piecewise linearly so that purely imaginary numbers with positive
imaginary part become yellow and purely imaginary numbers with negative
imaginary part become blue.
See~\cite{DLMF}*{\url{http://dlmf.nist.gov/help/vrml/aboutcolor\#S2.SS2}}.
For simplicity, we do not follow this approach here.)

Secondly, we add contour lines of complex numbers of the same phase at
integer multiples of $15$ degrees (see~\autoref{fig:domain-colouring-blackp}).
To that end, we change the intensity
of the colour by multiplying it with a sawtooth function
\[0.7 + (1.0 - 0.7) \cdot \left(\varphi / (\pi / 12) - \lfloor\varphi /
(\pi / 12)\rfloor\right).\]
Because phase corresponds to hue, the points of such a contour line are
all of the same colour.

Finally, we add contour lines of complex numbers of the same modulus on
a log-scale (see~\autoref{fig:domain-colouring-reference-image}). To
that end, we change the intensity of the colour by multiplying it with
a sawtooth function
\[0.7 + (1.0 - 0.7) \cdot \left(\log (r) / (\pi / 12) - \lfloor\log (r) /
(\pi / 12)\rfloor\right).\]
Note that the contour lines of phase and modulus intersect each other
orthogonally.
The scaling factor $1 / (\pi / 12)$ in the sawtooth function for the
modulus contour lines deliberately matches the scaling factor used in
the sawtooth function for the phase contour lines.
Consequently, the regions enclosed by the contour lines of phase and
modulus are squarish in appearance.

\subsection{Complex square root}
\label{sec:square-root-visualization}

Recall the construction of a Riemann surface of the complex
square root from \autoref{sec:mathematical-background} where we glued
together its two branches at a branch cut along the negative real axis.

\begin{figure}[ht]
\centering
\setcounter{subfigure}{0}
\begin{subfigure}[b]{0.49\linewidth}
\includegraphics[width=\linewidth]{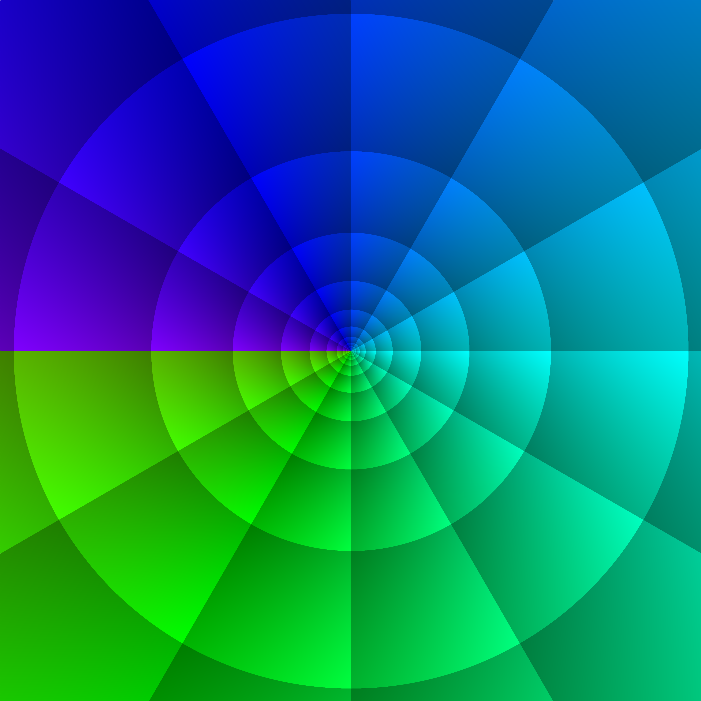}
\caption{}\label{fig:square-root-domain-colouring-negative}
\end{subfigure}
\begin{subfigure}[b]{0.49\linewidth}
\includegraphics[width=\linewidth]{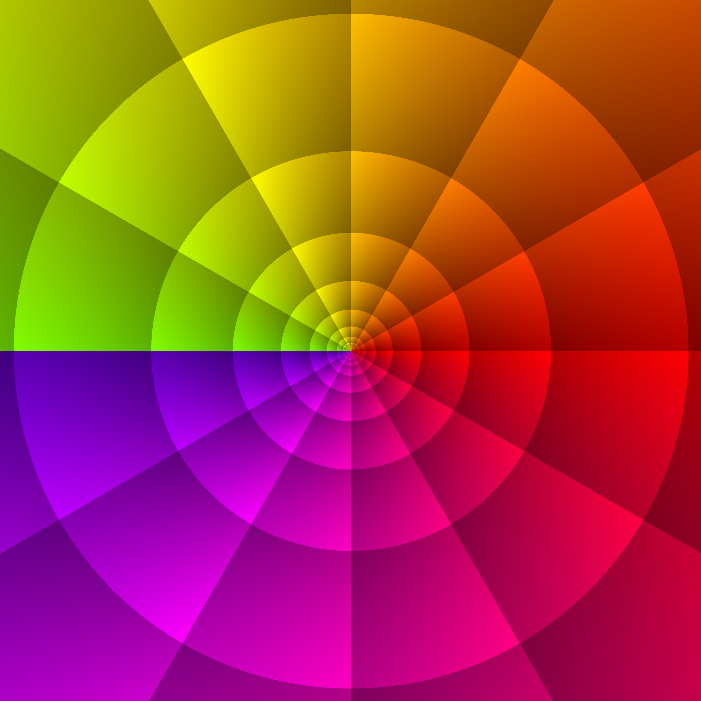}
\caption{}\label{fig:square-root-domain-colouring-positive}
\end{subfigure}
\caption{Domain colouring of the two sheets of the complex square root}
\label{fig:square-root-domain-colouring}
\end{figure}

The domain colouring of the two branches of the complex square root
over a square of side length $10$ centred at the origin is shown in
\autoref{fig:square-root-domain-colouring}.

On the sheet shown in~\autoref{fig:square-root-domain-colouring-negative},
the complex square root takes values with
negative real part (coloured green to blue). On the sheet shown
in~\autoref{fig:square-root-domain-colouring-positive},
it takes values with positive real part (coloured purple to yellow).
(The sheet shown in~\autoref{fig:square-root-domain-colouring-positive}
corresponds to the principal branch of the complex square root.)

On both sheets, twelve contour lines of phase are visible, half as many
as in the reference image. We can see that the phase of the complex square
root function changes at half the angular velocity of its argument.

Moreover, the discontinuity at the branch cut along the negative real axis is
clearly visible. We also see that there is a smooth transition between the
second (third) quadrant of
\autoref{fig:square-root-domain-colouring-negative} and the third (second)
quadrant of~\autoref{fig:square-root-domain-colouring-positive}.

\begin{figure}[ht]
\centering
\includegraphics[width=0.5\linewidth]{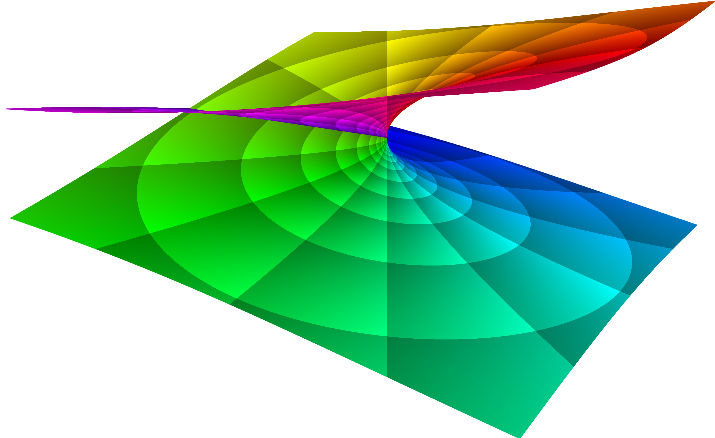}
\caption{Domain-coloured Riemann surface of the complex square root in
perspective projection}
\label{fig:square-root-perspective}
\end{figure}

\begin{figure}[ht]
\centering
\setcounter{subfigure}{0}
\begin{subfigure}[b]{0.24\linewidth}
\includegraphics[width=\linewidth]{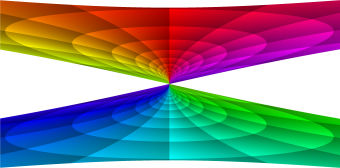}
\caption{}\label{fig:square-root-multiview-left}
\end{subfigure}
\begin{subfigure}[b]{0.24\linewidth}
\includegraphics[width=\linewidth]{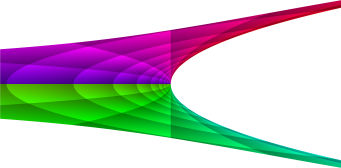}
\caption{}\label{fig:square-root-multiview-front}
\end{subfigure}
\begin{subfigure}[b]{0.24\linewidth}
\includegraphics[width=\linewidth]{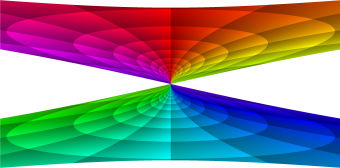}
\caption{}\label{fig:square-root-multiview-right}
\end{subfigure}
\begin{subfigure}[b]{0.24\linewidth}
\includegraphics[width=\linewidth]{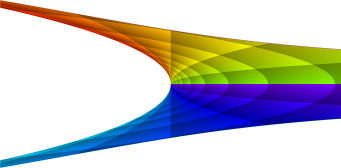}
\caption{}\label{fig:square-root-multiview-back}
\end{subfigure}
\caption{Domain-coloured Riemann surface of the complex square root
in orthogonal projection from left, front, right, and back (from left
to right)}
\label{fig:square-root-multiview}
\end{figure}

If we cut the two sheets along the negative real axis and glue the
upper side of the cut of one sheet to the lower side of the cut of the
other sheet, and vice versa, we obtain a Riemann surface of the complex
square root.
The resulting Riemann surface, produced with
\autoref{alg:riemann-surface-mesh} and~\autoref{alg:visualization}
using real part as height function, is shown in
\autoref{fig:square-root-perspective} (perspective) and
\autoref{fig:square-root-multiview} (multiview orthogonal).

Note that the self-intersection of the surface in
\autoref{fig:square-root-perspective} is only an artefact of using a
height function to map the Riemann surface mesh from real four-dimensional
to real three-dimensional space. Evidently, the two values of the complex
square root at each point of the self-intersection do not agree: they
are coloured differently, in green and purple, respectively.

In~\autoref{fig:square-root-multiview-front}, we see the
parabola that the real parts of the $y$-values describe according to the
equation $y^2 - x = 0$ when $x$ takes values on the non-negative real axis.

\subsection{Folium of Descartes}

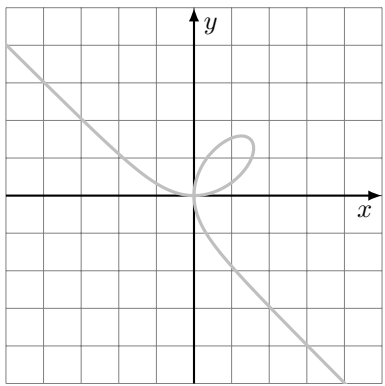
\begin{figure}[ht]
\centering
\begin{tikzpicture}[scale=0.5]
\draw[help lines] (-5,-5) grid (5,5);
\draw[thick,->, >=latex] (-5,0) -- (5,0) node[below left] {$x$};
\draw[thick,->, >=latex] (0,-5) -- (0,5) node[below right] {$y$};
\clip (-5,-5) rectangle (5,5);
\draw[very thick,domain=0:0.36,samples=100,lightgray] plot
({(6*\x*(-1+\x^2)^2)/(-1+3*\x^2+8*\x^3-3*\x^4+\x^6)},
{12*(\x^4-\x^2)/(3*\x^2+8*\x^3-3*\x^4+\x^6-1)});
\draw[very thick,domain=0:2.1, samples=100,lightgray] plot
({(-6*\x*(-1+\x^2)^2)/(-1+3*\x^2-8*\x^3-3*\x^4+\x^6)},
{12*(\x^4-\x^2)/(3*\x^2-8*\x^3-3*\x^4+\x^6-1)});
\end{tikzpicture}
\caption{The folium of Descartes as a real plane algebraic curve}
\label{fig:folium-real}
\end{figure}

\noindent The folium of Descartes is a classical plane algebraic curve
of order three,
\begin{equation}
\label{eq:folium}
\mathcal{C}\colon f (x, y) = x^3 + y^3 - 3 x y = 0.
\end{equation}
The cubic curve is nowadays called `folium' after the leaf-shaped
loop that it describes in the first quadrant of the real plane
(see~\autoref{fig:folium-real}).
It is named in honour of the French geometer René Descartes (1596--1650),
who was among the first mathematicians to introduce coordinates into
geometry.
Originally, the curve was called \emph{fleur de jasmin} since Descartes
and some of his contemporaries, who were working out the principles
of dealing with negative and infinite coordinates, initially wrongly
believed that the leaf-shaped loop repeated itself in the other quadrants
and therefore resembled a jasmine flower~\cite{Loria1910}*{p.~53}.

\begin{figure}[ht]
\centering
\setcounter{subfigure}{0}
\begin{subfigure}[b]{0.32\linewidth}
\includegraphics[width=\linewidth]{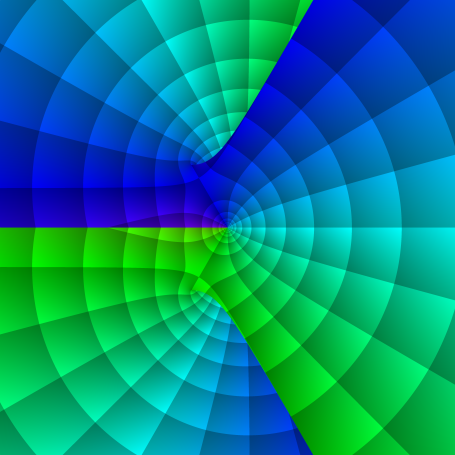}
\caption{}\label{fig:folium-domain-colouring-1}
\end{subfigure}
\begin{subfigure}[b]{0.32\linewidth}
\includegraphics[width=\linewidth]{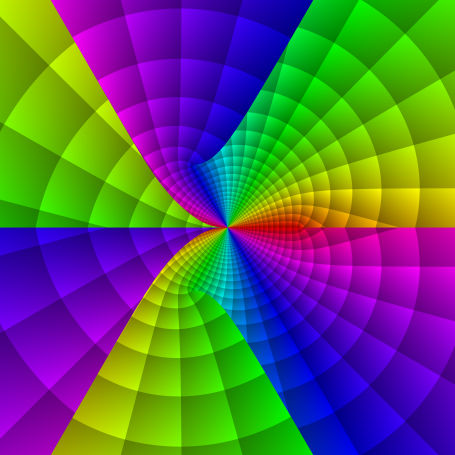}
\caption{}\label{fig:folium-domain-colouring-2}
\end{subfigure}
\begin{subfigure}[b]{0.32\linewidth}
\includegraphics[width=\linewidth]{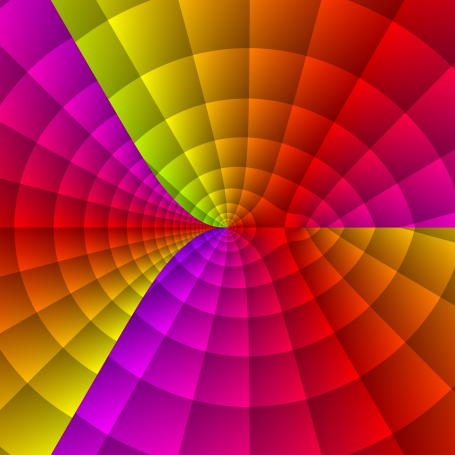}
\caption{}\label{fig:folium-domain-colouring-3}
\end{subfigure}
\caption{Domain colouring of three sheets of the folium of Descartes}
\label{fig:folium-domain-colouring}
\end{figure}

\noindent \autoref{fig:folium-domain-colouring} shows three
domain-coloured sheets of the folium of Descartes over a square of side
length $10$ centred at the origin of the complex plane. We can generate
these sheets by sorting the $y$-values that satisfy~\autoref{eq:folium}
at every point $x$ of the domain according to their real part. The sheet
shown in~\autoref{fig:folium-domain-colouring-1} uses the $y$-value
with the smallest real part, the sheet shown
in~\autoref{fig:folium-domain-colouring-2} the $y$-value with
the second-smallest real part, and the sheet shown
in~\autoref{fig:folium-domain-colouring-3} the $y$-value
with the largest real part.

We see that the first sheet carries $y$-values with negative real
part (coloured green to blue).
At the centre of the second sheet, we identify a zero of order two,
which we can recognize from the fact that the colours of the colour wheel
used in our reference image wind around it twice in the same order as
in the reference image. It is the node of the leaf-shaped loop.
The third sheet carries $y$-values with positive real part (coloured
purple to yellow).

There are three branch cuts (discontinuities of hue) on the first sheet,
six on the second sheet and three on the third sheets.
We can see how the sheets of the Riemann surface are connected to each
other along the branch cuts:
First and second sheet are connected at the branch cuts of the first
sheet. Second and third sheet are connected at the branch cuts of the
third sheet. First and third sheet are not connected directly with
each other.
(Imagine how much harder it would be to read this off from
\autoref{eq:folium}.)

Apart from the branch cuts, the map from $x$ to $y(x)$ is conformal
(angle-preserving) on every sheet. We can see that the contour lines
of phase and modulus intersect each other orthogonally on every sheet,
as in our reference image.

If we cut the sheets along the branch cuts and glue them together
correctly, we obtain a Riemann surface for the folium of Descartes.

\begin{figure}[ht]
\centering
\includegraphics[width=0.5\linewidth]{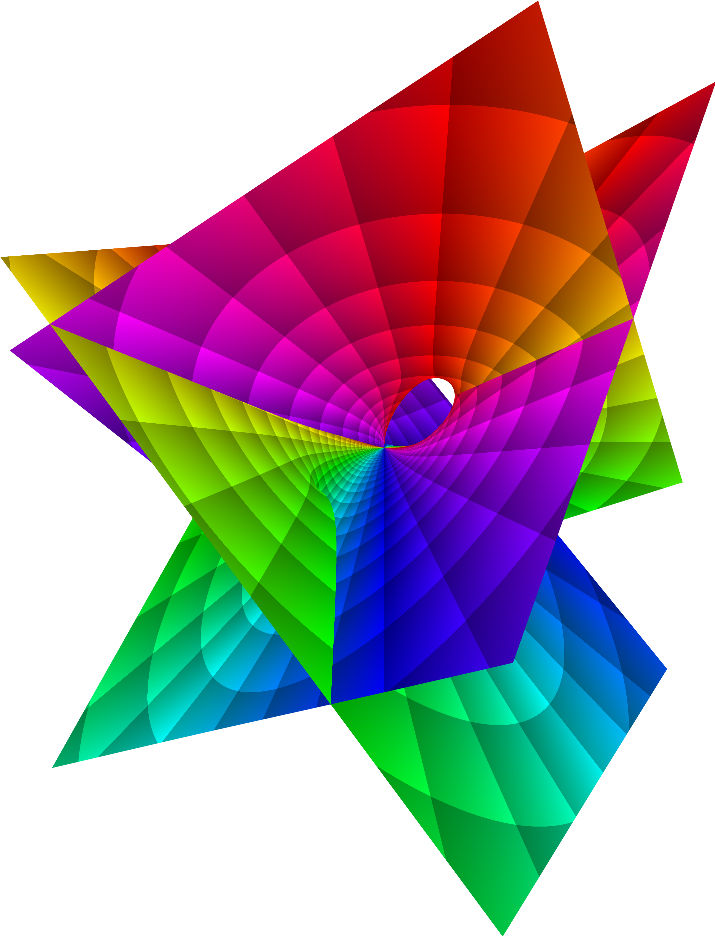}
\caption{Domain-coloured Riemann surface of the folium of Descartes in
perspective projection}
\label{fig:folium-perspective}
\end{figure}

\begin{figure}[ht]
\centering
\setcounter{subfigure}{0}
\begin{subfigure}[b]{0.24\linewidth}
\includegraphics[width=\linewidth]{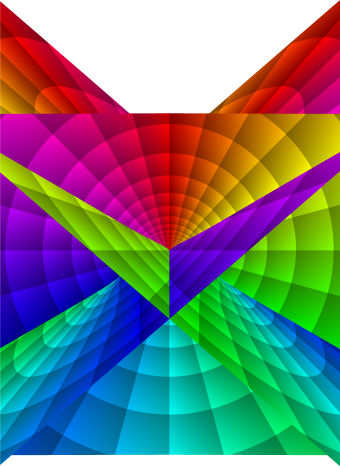}
\caption{}\label{fig:folium-multiview-left}
\end{subfigure}
\begin{subfigure}[b]{0.24\linewidth}
\includegraphics[width=\linewidth]{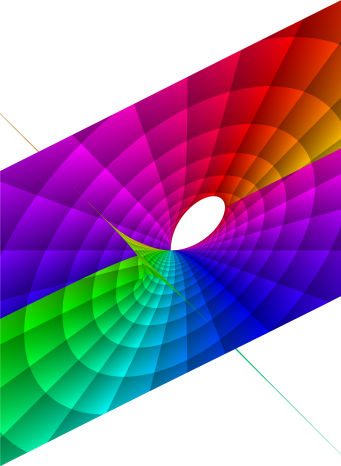}
\caption{}\label{fig:folium-multiview-front}
\end{subfigure}
\begin{subfigure}[b]{0.24\linewidth}
\includegraphics[width=\linewidth]{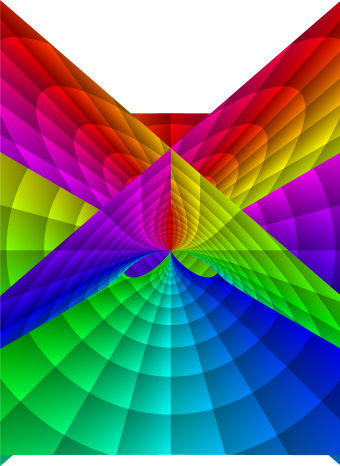}
\caption{}\label{fig:folium-multiview-right}
\end{subfigure}
\begin{subfigure}[b]{0.24\linewidth}
\includegraphics[width=\linewidth]{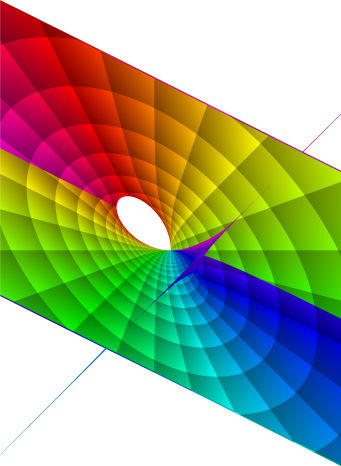}
\caption{}\label{fig:folium-multiview-back}
\end{subfigure}
\caption{Domain-coloured Riemann surface of the folium of Descartes in
orthogonal projection from left, front, right, and back (from left
to right)}
\label{fig:folium-multiview}
\end{figure}

The resulting Riemann surface, produced with
\autoref{alg:riemann-surface-mesh} and~\autoref{alg:visualization} using
real part as height function, is shown in~\autoref{fig:folium-perspective}
(perspective) and~\autoref{fig:folium-multiview} (multiview orthogonal).

Again, the self-intersections of the surface in
\autoref{fig:folium-perspective} are only an artefact of using a
height function to map the Riemann surface mesh from (real) four-dimensional
to (real) three-dimensional space.

\autoref{fig:folium-perspective} makes it obvious that cutting a Riemann
surface into sheets by sorting $y$-values by real part may be the most
straightforward but not necessarily the geometrically most appropriate
method.
Our Riemann surface of the folium of Descartes in large part appears to
be composed of three copies of the complex plane (which looks like our
reference image). Complications seem to arise only near the origin.

If we look closely at~\autoref{fig:folium-multiview-front},
we may see how we obtain the real folium of Descartes (as a real plane
algebraic curve) as the intersection of our Riemann surface mesh with
the $\operatorname{Re} x$-$\operatorname{Re} y$-plane. The leaf-shaped
loop is clearly visible as a hole in our visualization.
One of the `complex planes of which the Riemann surface is composed'
is so thin that it is barely visible from this perspective.
It is almost asymptotic to the `wings' of the folium of Descartes
(as a real plane algebraic curve) in the second and fourth quadrant of
the real $xy$-plane.

\begin{figure}[!ht]
\centering
\setcounter{subfigure}{0}
\begin{subfigure}[b]{0.49\linewidth}
\includegraphics[width=\linewidth]{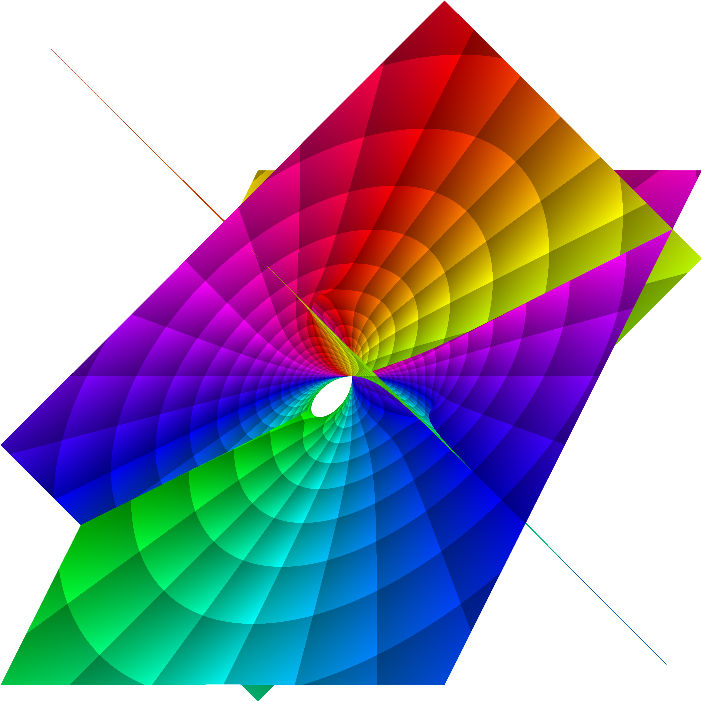}
\caption{}\label{fig:folium-green-loop}
\end{subfigure}
\begin{subfigure}[b]{0.49\linewidth}
\includegraphics[width=\linewidth]{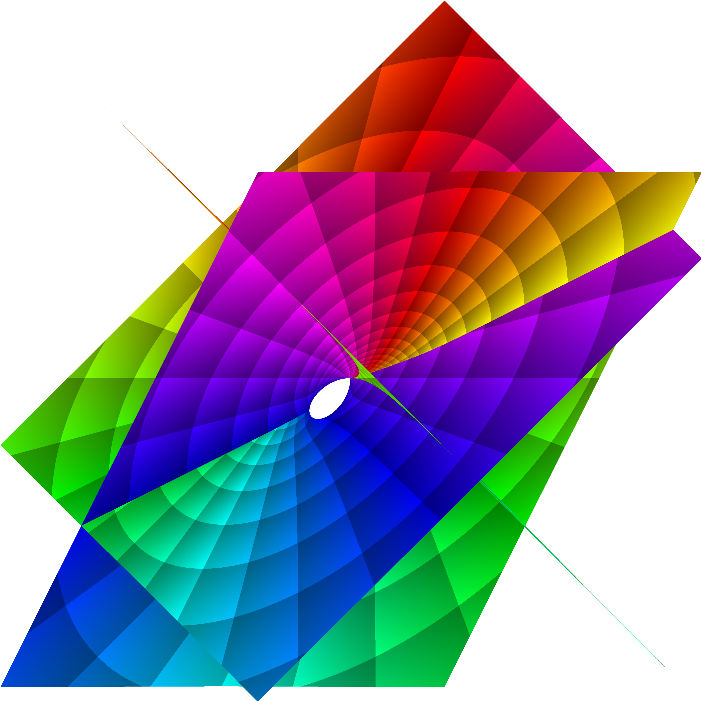}
\caption{}\label{fig:folium-blue-loop}
\end{subfigure}
\caption{Leaf-shaped loops of the folium of Descartes in complex directions.}
\label{fig:folium-more-leaves}
\end{figure}

Right below the centre of~\autoref{fig:folium-multiview-right}, we see
two leaf-shaped loops in complex directions.
Perhaps Descartes and his contemporaries were not entirely wrong after all to
believe that the folium of Descartes has more than one leaf.
Indeed, if we let $x' = \mathrm{e}^{\pm\mathrm{i} \pi / 3} x$, we discover
that in the $\operatorname{Re} x'$-$\operatorname{Re} y$-plane the curve
describes a leaf-shaped loop, which is exactly half as high as that in
the $\operatorname{Re} x$-$\operatorname{Re} y$-plane (this also
holds for the `wings') and rotated into a different quadrant
(see~\autoref{fig:folium-more-leaves}).

\section{Conclusion}

We have discussed algorithms for the generation of a Riemann surface mesh
of a plane algebraic curve (\autoref{alg:riemann-surface-mesh}) and its
visualization as a domain-coloured surface (\autoref{alg:visualization})
and their implementation using OpenGL and WebGL.
The WebGL implementation combines floating point textures, multiple render
targets, and a method due to \ocite{BoubekeurSchlick2008} to replace
the use of transform feedback and geometry shaders of the OpenGL implementation.
While the generation of the surface takes noticeable time in both
implementations, the visualization of a cached Riemann surface mesh is
possible with interactive performance.
This allows us to visually explore otherwise almost unimaginable
mathematical objects.
Sometimes the visualization makes properties of the plane algebraic
curves immediately apparent that may not so easily be read off from
its equation.
It is possible to turn these domain-coloured Riemann surface meshes into
physical models using a full colour 3D printer.

\section*{Funding}

This research was supported by DFG Collaborative Research Center TRR 109,
``Discretization in Geometry and Dynamics''.

\end{document}